# Controllable emergent spatial spin modulation in Sr$_2$IrO$_4$ by *in situ* shear strain

S. Pandey[1], H. Zhang[1*], J. Yang[1], A. F. May[2], J. Sanchez[3,4], Z. Liu[4], J.-H. Chu[4], J. W. Kim[3], P. J. Ryan[3,5], H. D. Zhou[1] and J. Liu[1*]

[1]Department of Physics and Astronomy, University of Tennessee; Knoxville, TN, USA, 37996.
[2]Materials Science and Technology Division, Oak Ridge National Laboratory; Oak Ridge, TN, USA, 37831.
[3]Advanced Photon Source, Argonne National Laboratory, Argonne, IL; USA, 60439.
[4]Department of Physics, University of Washington, Seattle, WA, USA; 98195.
[5]School of Physical Sciences, Dublin City University, Dublin 11; Ireland.

*corresponding author
**Email:** jianliu@utk.edu
hanzhang@cczu.edu.cn

## Abstract

Symmetric anisotropic interaction can be ferromagnetic and antiferromagnetic at the same time but for different crystallographic axes. We show that inducing competition of anisotropic interactions of orthogonal irreducible representations represents a general route to obtain new exotic magnetic states. We demonstrate it here by observing the emergence of a continuously tunable 12-layer spatial spin modulation when distorting the square-lattice planes in the quasi-2D antiferromagnetic Sr$_2$IrO$_4$ under *in situ* shear strain. This translation-symmetry-breaking phase is a result of an unusual strain-activated anisotropic interaction which is at the 4th order and competing with the inherent quadratic anisotropic interaction. Such a mechanism of competing anisotropy is distinct from that among the ferromagnetic, antiferromagnetic, and/or the Dzyaloshinskii–Moriya interactions, and it could be widely applicable and highly controllable in low-dimensional magnets.

Exotic spatial spin modulations, including various forms of spin spirals [1,2] and magnetic skyrmions [3-5], have been extensively studied for magnetoelectric effects [6], topology [7,8], spintronics [9-12], etc. Their common origin is competition between ferromagnetic, antiferromagnetic, and/or the Dzyaloshinskii–Moriya (antisymmetric anisotropic) interactions that favor parallel, antiparallel, and orthonormal spin arrangement, respectively. In contrast to these directional interactions, symmetric anisotropic interaction has an axial nature that it could be ferromagnetic for one axial component of the moments and antiferromagnetic for another at the same time, breaking rotational symmetry. Competition of anisotropic interactions of different symmetry channels could be an attractive alternative mechanism for stabilizing exotic magnetic phases. However, exploration of competing anisotropic interaction has been limited without efficient ways to introduce and tune the competition.

The strategy for realizing such competition could be remarkably simple and general. Since magnetic order often preserves certain rotational/mirror symmetry according to the inherent anisotropic interaction, one may apply shear strain to break the same symmetry to introduce antagonistic anisotropy. This necessarily causes competition of anisotropy with different irreducible point-group representations. Identifying materials for this method is however not trivial. The key ingredients are that the anisotropy must play a vital role in stabilizing the ordering and the magnetoelastic coupling of the shear-strain channel must be strong for inducing sizable anisotropy. As a pseudospin-half Mott insulator [13-27], $Sr_2IrO_4$ is an ideal system for this purpose. It is known for its analogy with the parent phase of high-$T_c$ cuprates [28-37], including the quasi-2D tetragonal lattice and an antiferromagnetic (AF) ordering below $T_N$~230K [14,16]. Due to the Ruddlesden-Popper crystal structure, the spontaneous quasi-2D AF order relies on the inter-plane anisotropic interaction to be stabilized, and breaks the $D_{4h}$ point group symmetry with the moments aligned along the $a$- or $b$-axis (Fig. 1a), i.e., orthorhombic twin domains of the $B_{1g}$ symmetry (Fig.1a) [38,39]. Because of the strong spin-orbit coupling, the magnetoelastic coupling can be described by the pseudo-Jahn-Teller effect, where the pseudospin-half wavefunction is highly sensitive to the lattice distortion. $Sr_2IrO_4$ thus provide a promising platform for exploring competing anisotropy under shear strain, i.e. strain of the $B_{2g}$ symmetry.

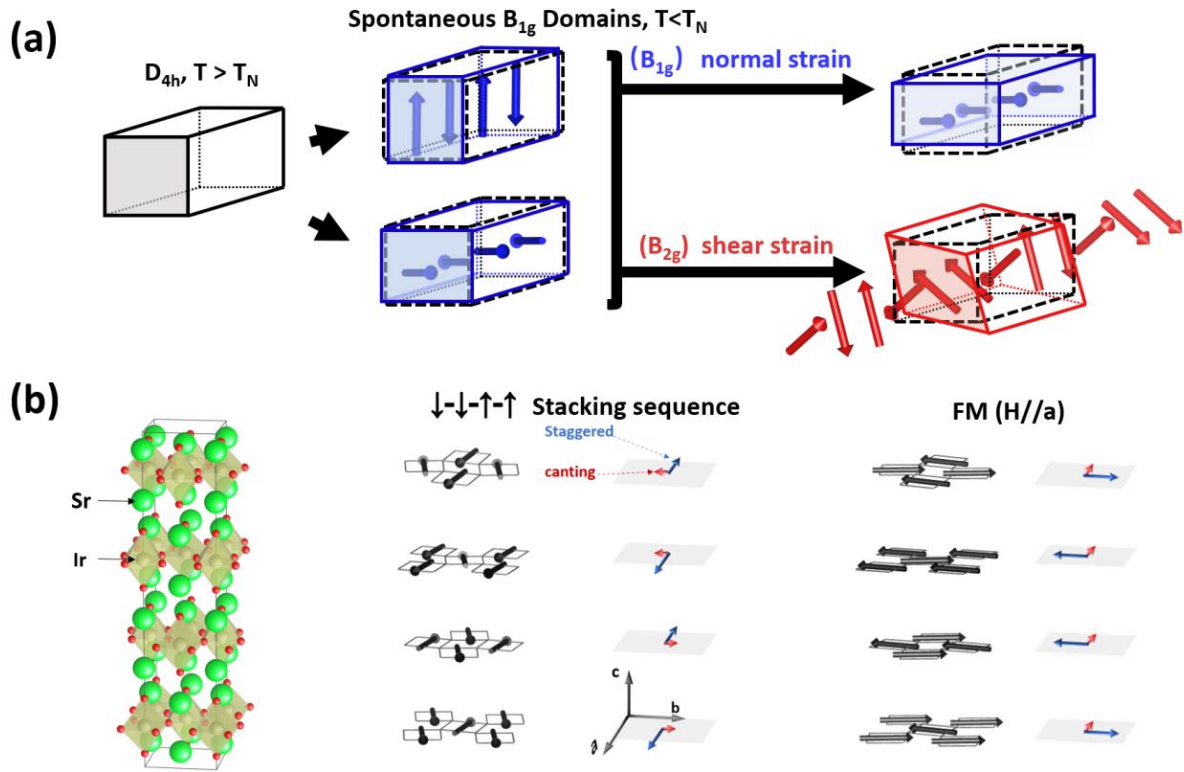

**Figure 1 Schematic of the $Sr_2IrO_4$ unit cell.** (a) schematic diagram showing symmetry-breaking of a tetragonal material by the magnetic order. The blue/red arrows denote the direction of the staggered moments in each $IrO_2$ plane in a tetragonal cell having 4 $IrO_2$-planes. (b) crystal structure of $Sr_2IrO_4$, with the ↓-↓-↑-↑ and FM stacking sequence below $T_N$ ~ 230 K. The net moment of each $IrO_2$ plane (red arrows) is locked with the staggered moment (blue arrows) with fixed chirality and proportionality in each layer.

Here we report that applying *in situ* shear strain to the *ab*-plane in $Sr_2IrO_4$ leads to a surprising translational symmetry-breaking modulation along the *c*-axis consisting of 12 $IrO_2$ planes (Fig.1a) that has not been observed before. Our study unambiguously shows that the spatial modulation is a result of competing anisotropy because the shearing turns on a hidden quartic anisotropic interaction of $B_{2g}$ symmetry that is *orthogonal* to the spontaneous quadratic anisotropy of $B_{1g}$ symmetry. Moreover, the competition and the modulation can be continuously and efficiently tuned here by applying a strain of less than 0.05%. Our observation of spatially modulated magnetic phases adds a significant representative to the expanding list of emergent phases due to tuning of interactions of different irreducible point-group representations such as electronic nematicity [40,41], quadrupolar order [42], and unconventional superconductivity [43-45].

As can be seen in Fig.1b, a key feature of $Sr_2IrO_4$ is the octahedral rotation around the *c*-axis [20], leading to a unit cell expansion from the original Ruddlesden-Popper structure. As a result, the spontaneous AF order fully preserves the translational symmetry and displays spin canting of the planar Ir AF moments within the *ab*-plane, which orthogonally locks the net moment and the staggered moment of each $IrO_2$ plane together with a fixed chirality and ratio [46-48]. Before discussing the shear strain effect, it is important to establish the normal strain effect for comparison. Since the spontaneous orthorhombic distortion is weak [39] and the system can be considered pseudo-tetragonal, applying normal strain, i.e., $B_{1g}$ strain, by external stress has been shown to be highly efficient in detwinning the AF domains [49,50] (Fig.1a). A more pronounced effect occurs when the strain is applied with a magnetic field. Since the inter-plane anisotropic interaction stabilizes a ↓-↓-↑-↑ (or ↓-↑-↑-↓) stacking sequence of the $IrO_2$ planes where ↑ and ↓ refer to the net moments [16,38], a sufficiently large field polarizes them through a metamagnetic phase transition [20] to a ↑-↑-↑-↑ sequence (Fig.1b), which for simplicity we call the ferromagnetic (FM) state hereafter. Figure 2a shows an example when we applied the field and the *in situ* $B_{1g}$ strain that stretched the *a*-axis and contracted the *b*-axis at 210 K by a piezo actuator, which delivers up to ~0.05% of anisotropic strain [49]. As can be seen, the transition monitored by magnetoconductance (MC) [51-54] is radically sharpened and shifted to a much lower field with H//*b*, whereas it is significantly broadened and shifted to much higher field with H//*a*. Such a strong anisotropic response with high tunability is a result of the uniaxial anisotropy induced by the $B_{1g}$ strain that restricts the metamagnetic transition to a sharp spin-flip or a broad continuous rotation depending on whether the field is parallel or perpendicular to the enforced easy axis [49,50].

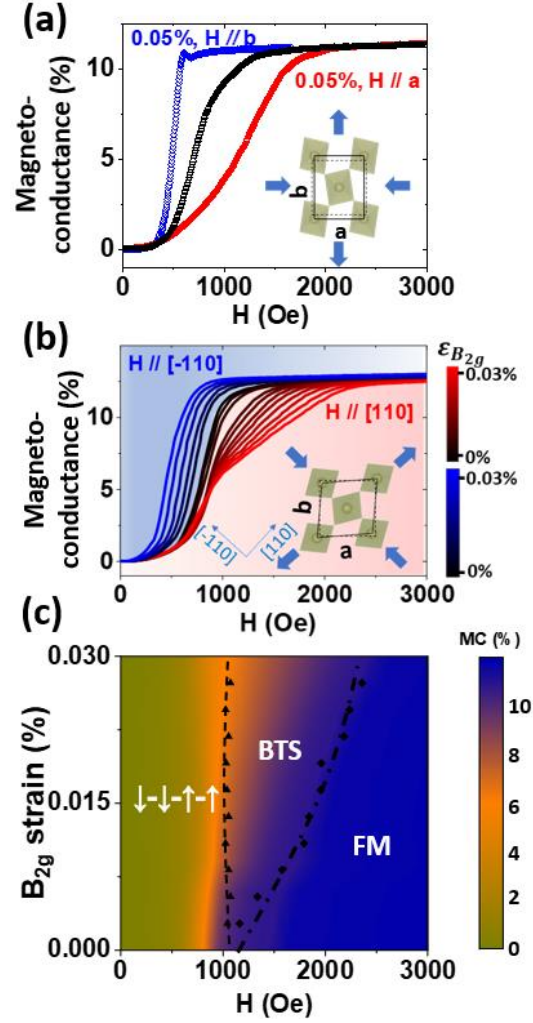

**Figure 2 Magnetoconductance under $B_{1g}$ and $B_{2g}$ strain.** (a) MC at 210 K without externally applied strain (black open triangles), under the applied $B_{1g}$ strain with H//a (red open triangles) and H//b (blue open squares). (b) MC at 210K under various $B_{2g}$ strains with H//[110] and with H//[-110]. Schematic diagrams of the distortions by the $B_{1g}$ and $B_{2g}$ strains are shown in the insets. The distortion is exaggerated for illustration purposes. (c) $B_{2g}$ strain-field phase diagram based on the MC results, where the phase boundaries are extracted from the kink positions in the MC curves.

The situation becomes much different under the shearing $B_{2g}$ strain (Fig.2b), which stretches the [110] axis and contracts the [-110] axis, i.e. $\varepsilon_{B_{2g}} \equiv (d_{110} - d_{-110})/(d_{110} + d_{-110})$ where $d_{110}$ and $d_{-110}$ represent the d-spacing of the (110) and (-110) planes, respectively. As shown in Fig.2b, while the transition with H//[-110] is also sharpened and shifted to lower field at 210 K similar to the case of H//b under $B_{1g}$ strain, the transition with H//[110] is clearly broken into two sections by an emerging kink at ~800 Oe. The section below the kink shows weak strain-dependence, whereas the section above is significantly extended to a higher field by increasing $\varepsilon_{B_{2g}}$, giving rise to a prolonged "*tail*" of an overall "*tied-up*" shape of the MC curves. An analysis of the derivative of the MC curve (see supplementary) reveals a fine movement of the kink that it is slightly upshifted first before being slightly downshifted with increasing $\varepsilon_{B_{2g}}$. A $B_{2g}$ strain-field phase diagram can now be constructed from the MC results as shown in Fig. 2c. A new phase is clearly emerging between the ↓-↓-↑-↑ state and the FM state with phase boundaries defined by the kink and the endpoint of the tail. This complex behavior is in sharp contrast with the case of H//*a* under $B_{1g}$ strain, suggesting a mechanism distinct from uniaxial anisotropy.

To resolve the underlying magnetic structure of the transition, we measured resonant x-ray magnetic scattering (RXMS) of the (10L) and (-10L) magnetic peaks at the Ir $L_3$-edge under *in situ* $B_{2g}$ strain with H//[-110] and H//[110], respectively. With $\varepsilon_{B_{2g}} = 0.015\%$ at 210 K and zero field, we observed the magnetic reflections at L = 20 and 22 associated with the two spontaneous $B_{1g}$ twin domains of the ↓-↓-↑-↑ state, respectively [49]. The presence of both domains without detwinning is consistent with the $B_{2g}$ symmetry of the applied strain. As shown in Fig.3a, when H//[-110] is turned on to induce the metamagnetic transition, the L = 21 peak associated with FM order parameter of the net moments emerges and increases rapidly at the expense of the L = 20 and 22 peaks until both of them vanish[*]. The overall conversion between the ↓-↓-↑-↑ state and the FM state is similar to that in the strain-free sample [55], despite the downshifting of the transition to the region of ~500 - 700 Oe, consistent with the MC data.

---

[*] The L = 20 peak vanishes a bit earlier than the L = 22 peak due to a small misalignment of the field that suppresses one domain slightly faster than the other.

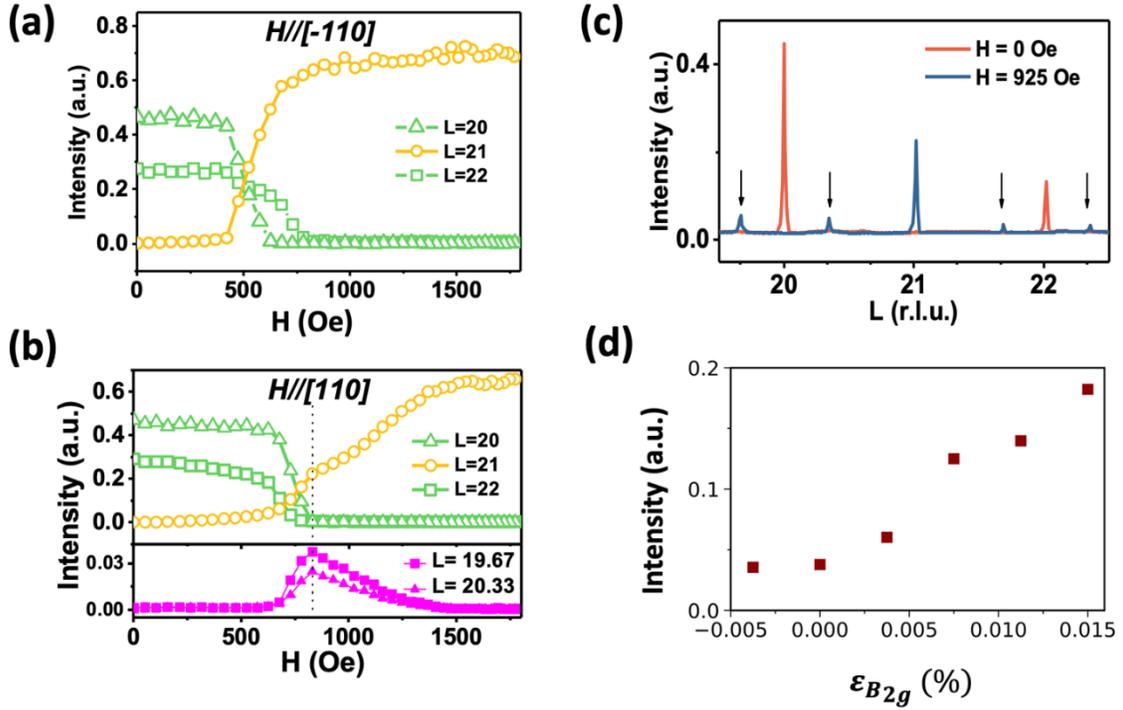

**Figure 3 Resonant x-ray magnetic scattering.** (a) Evolution of the (1 0 20-22) magnetic reflections associated with the two twin domains of the ↓-↓-↑-↑ state and the FM state as functions of the external magnetic field when H//[-110] and $\varepsilon_{B_{2g}} = 0.015\%$. (b) Evolution of the equivalent (-1 0 20-22) magnetic reflections when H//[110] and $\varepsilon_{B_{2g}} = 0.015\%$ is shown on the top panel. The bottom panel shows the evolution of (-1 0 $19\frac{2}{3}$) and (-1 0 $20\frac{1}{3}$) peaks. (c) L-scans of the (-1 0 L) rod when H//[110] and $\varepsilon_{B_{2g}} = 0.015\%$. The navy curve taken at H=925 Oe shows the emergence of 1/3 satellite peaks highlighted by the black arrows. The red curve taken at zero field is shown for comparison. (d) The emergence and increase of the (-1 0 $19\frac{2}{3}$) peak intensity as a function of the varying strain from -0.004% to 0.015%.

The transition process turns out very differently with H//[110] as shown in Fig.3b. When both L = 20 and 22 peaks drop to zero at 825 Oe, the L = 21 peak only gains a partial increase that is far from its saturated value. Most of the intensity is instead from a slow increase from 825 to 1500 Oe. The slope change at 825 Oe leads to a clear kink and an overall shape that highly resembles the MC curve, suggesting that the extended region above the kink is neither the ↓-↓-↑-

↑ state nor the FM state but a distinct new state with a finite FM component. After performing a survey at other L positions at 925 Oe such as that shown in Fig.3c, we discovered emerging satellite magnetic peaks at L = $2n \pm \frac{1}{3}$, indicating translational symmetry-breaking along the *c*-axis with a tripled size. Such a 12-$IrO_2$-plane modulation of the staggered moment corresponds to a 6-$IrO_2$-plane modulation of the net moment according to the out-of-phase octahedral rotation within the structural cell (Fig.1b). We denote this state as the BTS state hereafter. The field-dependence of the satellite peak intensities displayed in Fig.3b exhibits a clear correlation with the integer peaks. It is now apparent that the ↓-↓-↑-↑ state first transforms into the BTS state, and the FM component of the BTS state gradually increases at the expense of the amplitude of the spatial modulation until reaching the FM state. Figure 3d further shows the increase of the satellite peak intensity at 925 Oe as a function of $\varepsilon_{B_{2g}}$, demonstrating that the $B_{2g}$ strain not only stabilizes the spatial modulation but also efficiently tunes its amplitude. These results confirm the new phase in the phase diagram to be the BTS state.

The emergence of the BTS state corroborates that the $B_{2g}$ strain effect is to compete with the spontaneous anisotropy that favors $B_{1g}$ symmetry rather than changing the easy axis. To unveil this competition, we investigated the minimal free energy model that captures the BTS state. In virtue of the robust intra-plane AF order and the locking between the net moment and staggered moment [47], the stacking of the quasi-2D planes can be modeled as an effective 1D system of the net moments $\boldsymbol{M}_j = M\boldsymbol{S}_j$, where $M$ is the moment size and $\boldsymbol{S}_j$ is the unit vector in the *ab*-plane characterized by angle $\theta_j$ with respect to the *a*-axis. The ↓-↓-↑-↑ state, BTS state, and FM state are thus characterized by a 1D wave vector $q$ = 1/4, 1/6, and 0, respectively. The free energy can be written as:

$$F\left(\boldsymbol{M}_j\right) = E_0 + E_H + E_{B_{1g}} + E_{B_{2g}} \tag{1}$$

where $E_0$ includes inherent interactions that account for the spontaneous ↓-↓-↑-↑ state of $B_{1g}$ symmetry, $E_H$ is the Zeeman energy, $E_{B_{1g}}$ and $E_{B_{2g}}$ represent interactions induced by the externally applied strain of $B_{1g}$ and $B_{2g}$ symmetry, respectively. At a fixed temperature below $T_N$, $M$ is considered as a constant under the perturbations of field and strain, and it is ~0.04 $\mu_B$ at 210 K according to magnetization. Thus, all $\theta_j$-independent terms are neglected, and the ground state is obtained by energy minimization with respect to $\theta_j$. By mapping the reported inter-plane anisotropic interactions and the pseudospin-lattice coupling of Sr$_2$IrO$_4$ [39,49] to this effective model, one has

$$E_0 = \delta_c \sum_j (-1)^j (S_j^a S_{j+1}^a - S_j^b S_{j+1}^b) - k_b \sum_j Q_j^{(1)^2} - k_b' \sum_j Q_j^{(1)} Q_{j+1}^{(1)} \qquad (2)$$

Where $Q_j^{(1)} \equiv S_j^{a2} - S_j^{b2}$ is the quadrupole of B$_{1g}$ symmetry. The $\delta_c$-term is the anisotropic exchange interaction of B$_{1g}$ symmetry with an alternating sign that favors a $q = 1/4$ modulation. It becomes the ↓-↓-↑-↑ state with the $k_b$-term and the $k_b'$-term, which have the A$_{1g}$ symmetry and force the moments along the $a$- or $b$-axis [38,39] (see supplementary). We constrain $\delta_c$, $k_b$, and $k_b'$ by reproducing the metamagnetic transition at zero strain and under the B$_{1g}$ strain by including a uniaxial anisotropy term in $E_{B_{1g}}$ [49] (Fig. S3).

Next we turn off $E_{B_{1g}}$ and proceed to simulate the B$_{2g}$ strain effect. We first tried B$_{2g}$ symmetry terms of uniaxial and exchange anisotropy. But both fail to describe the experimental results (see supplementary). This inconsistency leads us to consider increasing the order of the anisotropic interaction from quadratic to quartic. The only nontrivial quartic term of B$_{2g}$ symmetry has the form (see Methods)

$$E_{B_{2g}} = \varepsilon_{B_{2g}} \kappa \sum_j \mathbf{S}_j \cdot \mathbf{S}_{j+1} (Q_j^{(2)} + Q_{j+1}^{(2)}) \qquad (3)$$

or equivalently

$$E_{B_{2g}} = \varepsilon_{B_{2g}} \kappa \sum_j \mathbf{S}_j \cdot (\mathbf{S}_{j-1} + \mathbf{S}_{j+1}) Q_j^{(2)} \qquad (4)$$

where $Q_j^{(2)} \equiv S_j^a S_j^b$ is the quadrupole of B$_{2g}$ symmetry and $\kappa$ is the coupling coefficient. Figure 4a shows the calculated metamagnetic transition, which is characterized by the $q = 0$ order parameter and turns out to capture the observed characteristics remarkably well, especially the two-section shape of the transition with an extended tail region when H//[110]. Even the fine movement of the kink in the MC curve is mimicked by the strain dependence of the critical field that separates the two sections. More importantly, this quartic anisotropy correctly predicts a finite $q = 1/6$ modulation within the tail region where the $q = 1/6$ modulation appears and slowly decreases with increasing field.

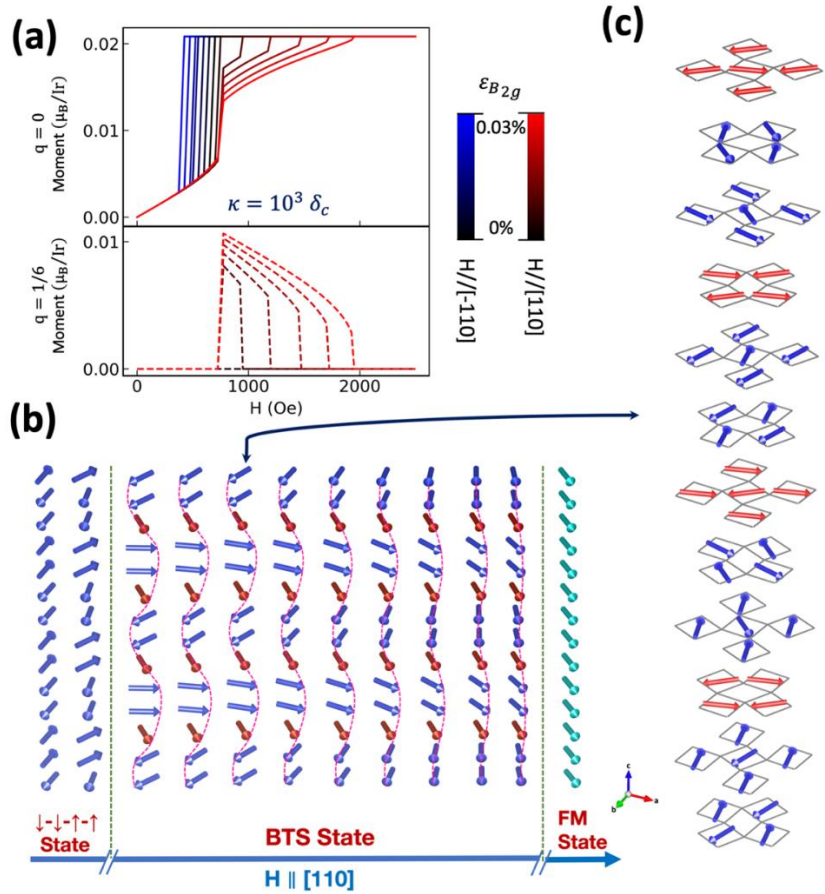

**Figure 4 Simulation of the free-energy model.** (**a**) Simulation of metamagnetic transition when including the $\kappa$ term that is coupled to $B_{2g}$ strain. The solid lines represent the order parameter of the $q=0$ (FM) state while the dashed line is for the $q=1/6$ (BTS) state. The coupling coefficient $\kappa = 10^3\, \delta_c$. (**b**) Schematic diagram showing the net moment evolution during the ↓↓↑↑-BTS-FM metamagnetic phase transition. The full magnetic structure in a 12-IrO$_2$-layer cell is shown in (**c**) for one of the BTS configurations.

Figure 4b shows a representative evolution of the simulated spatial modulation. Initially, the moments cant away from the *a*/*b*-axis towards [110] to gain Zeeman energy while staying in pairs to maintain the ↓-↓-↑-↑ state. When the field increases further, the $q = 1/6$ modulation sets in by "*inserting*" an unpaired site in between the adjacent pairs and breaking the translation symmetry. The modulation amplitude of this BTS state then slowly decreases with increasing field until both the unpaired and paired moments become fully aligned. The periodic emergence of the unpaired sites is evidently the key since it plays a domain wall-like role in separating the pairs, allowing the $\kappa$-term to save more energies on the paired sites while letting the $\delta_c$-interaction to concentrate more on the unpaired site. The full 12-IrO$_2$-layer magnetic structure is also shown for one of the BTS configurations as an example to illustrate the modulation in Fig.4c.

Given the excellent agreement with the experiment, we comment on the important differences of the quartic anisotropic interaction from the usual quadratic ones, and why it enables the emergent BTS state. Specifically, increasing the order of anisotropic interaction necessarily allows more intermediate angular configurations with no energy gain or loss, i.e., nodes or saddle points in the energy landscape (supplementary). This can be understood from Eq.(4), where the nearest-neighbor interaction is *gated* by $Q_j^{(2)}$, controlling the sign of the interaction to be positive or negative. It could even be zero whenever $\boldsymbol{S}_j$ is along the *a/b*-axis, and the interaction would be zero in this case regardless of the direction of $\boldsymbol{S}_{j-1}$ and $\boldsymbol{S}_{j+1}$. This character is absent in quadratic interactions and provides more options to find compromised configurations when competing with the $\delta_c$-interactions of orthogonal symmetry. We further confirmed this anisotropy competition by simulating a hypothetical situation where the A$_{1g}$ $k_b$ and $k_b'$ terms are removed (see supplement).

Overall, we demonstrate that introducing anisotropy that is orthogonal to the spontaneous one could lead to translational symmetry-breaking and spatial spin modulation. The possibility of introducing the competition between the B$_{1g}$ and B$_{2g}$ anisotropy in Sr$_2$IrO$_4$ by *in situ* shear strain allows continuous tuning of the spatial modulation. This approach to trigger competition and obtain controllable new phases could be applicable to many quasi-2D materials due to the critical role of anisotropy in stabilizing low-dimensional magnetic orders. The emergence of the B$_{2g}$ quartic interaction also highlights the rich unexplored physics of higher-order anisotropy. To our knowledge, quartic anisotropic interaction has not been observed before. Since quartic exchange interactions in Mott insulators arise from the fourth-order perturbation [56], distinct hopping processes must be involved at the same time, such as that between the planes and within the planes. These results should motivate theoretical investigations on extracting the underlying microscopic model as well as exploration of high-order anisotropic interactions in other materials.

**Acknowledgment:**

Sample synthesis (A.F.M.) was supported by the U. S. Department of Energy, Office of Science, Basic Energy Sciences, Materials Sciences and Engineering Division. The in-situ strain control and measurement setup are partially supported by AFOSR DURIP Award FA9550-19-1-0180 and as part of Programmable Quantum Materials, an Energy Frontier Research Center funded by the U.S. Department of Energy (DOE), Office of Science, Basic Energy Sciences (BES), under award DE-SC0019443. J.H.C. acknowledge the support of the David and Lucile Packard Foundation. Transport measurement and modeling analysis are supported by the U. S. Department of Energy under grant No. DE-SC0020254. This research used resources of the Advanced Photon Source, a U.S. Department of Energy (DOE) Office of Science User Facility operated for the DOE Office of Science by Argonne National Laboratory under Contract No. DE-AC02-06CH11357. The authors thank Cristian Batista for valuable discussions; and Randal R. McMillan and Hao Zhang for providing technical support.